# All-optical coherent control of vacuum Rabi oscillations

Ranojoy Bose[1,†], Tao Cai[1,†], Kaushik Roy Choudhury[1], Glenn S. Solomon[2] and Edo Waks[1,2,*]

[1]Department of Electrical Engineering and Institute for Research in Electronics and Applied Physics, University of Maryland, College Park, MD 20742.

[2]Joint Quantum Institute, University of Maryland and National Institute of Standards and Technology, College Park, MD 20742

[†]These authors contributed equally

[*]Correspondence should be sent to E.W.: edowaks@umd.edu

**When an atom strongly couples to a cavity, it can undergo coherent vacuum Rabi oscillations. Controlling these oscillatory dynamics quickly relative to the vacuum Rabi frequency enables remarkable capabilities such as Fock state generation and deterministic synthesis of quantum states of light, as demonstrated using microwave frequency devices[1,2]. At optical frequencies, however, dynamical control of single-atom vacuum Rabi oscillations remains challenging. Here, we demonstrate coherent transfer of optical frequency excitation between a single quantum dot and a cavity by controlling vacuum Rabi oscillations. We utilize a photonic molecule[3-7] to simultaneously attain strong coupling and a cavity-enhanced AC Stark shift. The Stark shift modulates the detuning between the two systems on picosecond timescales, faster than the vacuum Rabi frequency. We demonstrate the ability to add and remove excitation from the cavity, and perform coherent control of light-matter states. These results enable ultra-fast control of atom-cavity interactions in a nanophotonic device platform.**

Much of the prior work investigating atomic systems strongly coupled to optical cavities has operated in the static regime. In this regime the coupling between the two systems remains constant and the signature of strong coupling is observed either in the frequency domain in the form of vacuum Rabi splitting[8-11], or by direct time-domain observation of vacuum Rabi oscillations[12]. Recently, there has been significant experimental progress in optically controlling the dynamical response of atomic systems strongly coupled to cavities for applications such as optical switching[13-16], reversible storage of photonic qubits[17,18], and hybrid quantum information processing[19,20]. These works have all operated in the adiabatic regime where the duration of the optical control pulse was long compared to the vacuum Rabi frequency.

When the interaction between the atom and cavity is modulated fast compared the vacuum Rabi frequency, the system undergoes diabatic rapid passage. In this regime it becomes possible to coherently transfer energy between an atomic excitation and a cavity photon by directly controlling vacuum Rabi oscillations. This coherent transfer has been effectively implemented at microwave frequencies and has enabled capabilities such as Fock state generation[1] and synthesis of arbitrary photonic wavefunctions[2]. At optical frequencies, however, diabatic control of vacuum Rabi oscillations between a single atomic system and a cavity remains difficult.

In this letter we report a demonstration of controlled transfer of excitation between a semiconductor quantum dot and a strongly coupled optical cavity by directly controlling vacuum Rabi oscillations diabatically. We use a pulsed AC Stark shift to control the detuning between the quantum dot and cavity on picosecond timescales, enabling ultra-fast control of light-matter interactions. We demonstrate transfer of excitation between the cavity and quantum dot, and show that this process is coherent. Furthermore, we demonstrate that this method enables coherent control of light-matter states, an important building block for synthesizing photon wavefunctions. These results could ultimately enable controlled generation of quantum states of light in the optical domain at gigahertz rates[21].

To understand how a Stark shift enables coherent control of vacuum Rabi oscillations, we first consider a simplified model of the system described by the level structure shown in Figure 1a. We restrict our attention to the first excitation manifold, a valid approximation when the system is weakly excited. We treat the quantum dot as a two-level atomic system and adopt the notation $\left| G,n \right\rangle$ and $\left| E,n \right\rangle$ to denote quantum states where the atom is in its ground and excited state respectively, and the cavity contains $n$ photons. In the figure, $\Delta_c$ is the detuning between the atom and cavity, and $g$ is the cavity-quantum dot coupling strength. An off-resonant optical pulse, expressed as a classical Rabi frequency $\Omega(t)$, excites the quantum dot with a detuning $\Delta$. The optical pulse induces an AC Stark shift that controls the detuning between the dot and cavity. We assume $\Delta \gg \Omega(t)$ for all time so that the off-resonant pulse only interacts with the quantum dot through a virtual transition. In this limit, the atom-cavity system evolves according to the following effective Hamiltonian (see Supplement):

$$\mathbf{H}_{eff} = \hbar\Delta_{nl}\sigma_+\sigma_- + \hbar g\left(\hat{\mathbf{a}}^\dagger\sigma_- + \sigma_+\hat{\mathbf{a}}\right) \tag{1}$$

where $\hat{\mathbf{a}}$ is the bosonic annihilation operator for the cavity mode, $\sigma_+$ and $\sigma_-$ are the quantum dot dipole raising and lowering operators, and $\Delta_{nl} = \Delta_c - \dfrac{2\Omega^2(t)}{\Delta}$ is the cavity-quantum dot detuning in the presence of an AC Stark shift.

For the Hamiltonian described in equation (1), states $|E,0\rangle$ and $|G,1\rangle$ comprise a two-level system that can be coherently controlled by the Stark field $\Omega(t)$ that rapidly modulates their relative detuning. The effect of the Stark field is particularly simple to derive in the ideal limit of diabatic rapid passage, where the field is turned on and off instantaneously. For the special case where only the quantum dot is excited and the Stark shift tunes it onto cavity resonance, the probability that the system occupies the state $|G,1\rangle$ after the Stark pulse is $P_{G,1} = \sin^2\left(g\tau\right)$, where $\tau$ is the duration of the Stark pulse (see Supplement). Thus, the quantum dot excitation coherently transfers to the cavity, with perfect transfer occurring at the condition $2g\tau = \pi + m2\pi$ where $m$ is an integer. We obtain the same results for the quantum dot excitation when the cavity is initially excited, indicating that this process and can be used to transfer excitation to either system.

Achieving a large AC Stark shift using the excitation scheme illustrated in Figure 1a is challenging because the field $\Omega(t)$ drives the cavity mode off-resonance. Therefore, the majority of the field reflects and only a small amount of power drives the quantum dot. Recently, we have demonstrated that a photonic molecule can solve this problem[5]. A photonic molecule is composed of two cavities coupled by a fast photon tunneling interaction[3-7]. These photonic structures exhibit two non-degenerate modes, one that strongly couples to the quantum dot and a second that can induce a cavity-enhanced AC Stark effect.

We utilize a photonic crystal implementation of a photonic molecule in order to implement controlled transfer. Figure 1b shows a scanning electron micrograph (SEM) image of the fabricated device composed of two evanescently coupled photonic crystal cavities, along with the calculated modes (see Methods). The photonic molecule exhibits two coupled modes composed of symmetric and anti-symmetric combinations of the individual cavity modes (top and bottom respectively in Figure 1b).

We characterize the response of the fabricated device under continuous wave excitation using a broadband LED that acts as a white light source (see Methods). Figure 1c shows the reflection spectrum of the device, taken at 45 K. The spectrum exhibits two peaks corresponding to the coupled cavity modes, denoted M1 (at $\omega_-$= 925.84 nm) and M2 (at $\omega_+$= 927.48 nm), which are spectrally separated by 565 GHz. The spectrum also shows an additional peak corresponding to a quantum dot resonance that is red-detuned from mode M1 (labeled QD in the figure). We note that another quantum dot can be seen to couple to mode M2, but is far detuned from mode M1 and plays no role in our measurements. The quality factor ($Q$) of modes M1 and M2 are calculated from the spectrum to be 18,500 and 12,000 respectively, corresponding to energy decay rates of $\kappa_1/2\pi$=17.7 GHz and $\kappa_2/2\pi$=28.3 GHz. Second order correlation measurements of the quantum dot emission reveal a clear anti-bunching, ensuring that we are working with a single dot (see Supplement).

Figure 1d shows the cavity reflection spectrum near the resonance of mode M1 as a function of temperature. The quantum dot resonance red-shifts with increasing temperature and exhibits an anti-crossing when tuned across the cavity resonance. We calculate a coupling strength of $g/2\pi = 8.1$ GHz by fitting the reflection spectrum when the quantum dot is resonant with the cavity to the theoretically predicted spectrum (see Supplement). The coupling strength satisfies the condition $4g > \kappa_1$ indicating that we are operating in the strong coupling regime[8-11]. We note that the vacuum Rabi period, given by $\tau_r = \frac{\pi}{\sqrt{g^2-(\kappa_1/4)^2}} = 74$ ps in the presence of cavity losses[9,10], is much longer than the cavity lifetime of mode M2 given by $\tau_2 = \frac{1}{\kappa_2} = 6$ ps. Thus, we can dynamically excite mode M2 to induce a Stark shift on timescales that are much shorter than $\tau_r$.

To show that we can coherently control the strongly coupled system on fast timescales, we utilize picosecond optical pulses both for excitation and to induce a Stark shift (see Methods). We first investigate energy transfer from the quantum dot to the cavity. We set the detuning between the two systems to $\Delta_c = 18$ GHz, which is sufficiently large to ensure that the stationary states of the system are well-approximated by the bare quantum dot and cavity excitations. An excitation pulse drives the quantum dot resonance, and a subsequent Stark pulse drives the resonance of mode M2. The Stark pulse induces a pulsed AC Stark shift that transfers energy to the cavity on timescales that are short compared to the decay rate of the quantum dot exciton. We then allow the system to radiatively decay and measure the total emission spectrum to determine the fraction of light emitted at the cavity frequency.

Figure 2a shows the measured cavity emission spectrum as a function of $\Delta\tau$, the time delay between the excitation and Stark pulse. For $\Delta\tau < 0$, the cavity emits a constant background independent of the Stark pulse delay. This background is due to partial spectral overlap between the excitation pulse and cavity resonance, as well as non-resonant energy transfer of the quantum dot excitation[22]. However, when $\Delta\tau > 0$ we observe enhanced emission at the cavity resonance indicating a transfer of excitation from the quantum dot. Figure 2b plots the intensity at the cavity emission resonance ($\Delta\lambda=0$) as a function of $\Delta\tau$. The cavity emission exhibits an oscillatory behavior as a result of perturbed free induction decay which occurs when a quantum dot undergoes a transient energy shift[23-26]. The emission decays to the background level with a decay time of 56 ps (determined by fitting the data to an exponentially decaying sinusoid) due to relaxation rate of the quantum dot.

In order to show that energy can also be transferred from the cavity to the quantum dot, we perform a second experiment where we tune the excitation pulse to the cavity resonance. In Figure 2c we show the emission spectrum as a function of $\Delta\tau$, while Figure 2d plots the intensity at the quantum dot emission resonance. The quantum dot intensity exhibits a sharp enhancement near $\Delta\tau = 0$ due to transfer of excitation from the cavity. The transfer occurs over a time window of only 14 ps (at full width half maximum) because the cavity exhibits a faster decay rate than the quantum dot.

We can compare the experimental results to the theoretically predicted behavior of the system. Our previous analysis based on the Hamiltonian in equation (1) used a simplified theoretical model that assumed ideal diabatic rapid passage, ignored damping of both the cavity field and quantum dot exciton, and did not account for finite excitation time of the system. In order to perform more realistic calculations, we numerically solve the master equation[27] which rigorously accounts for cavity field decay and exciton dephasing, and can calculate the system response for laser pulses of arbitrary temporal shapes (see Supplement). Figures 2e-f plot the calculated spectrum and cavity resonance intensity as a function of $\Delta\tau$ for the experimental conditions used to attain the results of Figures 2a-b. Calculations exhibit good agreement with the measured spectrum and predict the observed transfer near $\Delta\tau = 0$ as well as oscillations due to perturbed free induction decay. Figures 2g-h plot the calculated spectrum and emision intensity for the experimental conditions used to obtain Figures 2c-d, and also reproduce the expected transfer of energy sharply peaked near $\Delta\tau = 0$.

We next investigate the energy transfer as a function of the Stark laser power. We excite the quantum dot resonance and fix $\Delta\tau$ to the peak transfer point observed in Figure 2b. Figure 3a plots the measured spectrum as a function of average Stark field power (determined after the focusing objective). The spectrum shows out-of-phase oscillations between the quantum dot and cavity emission. These oscillations become clearer when we plot the intensity at the resonant frequency of the cavity ($\Delta\lambda = 0$) and quantum dot ($\Delta\lambda = 0.07$ nm) as a function of Stark field power, as shown in Fig. 3b. The origin of the oscillations can be understood using the simplified model described by the Hamiltonian in equation (1). From this model, we can show that $P_{G,1} \propto \sin^2(\omega_r\tau/2)$, where $\omega_r = \sqrt{(2\Omega^2/\Delta - \Delta_c)^2 + 4g^2}$ is the vacuum Rabi frequency of the Stark shifted quantum dot and $\Omega$ is the amplitude of the Stark pulse, expressed as a classical Rabi frequency. By increasing Stark pulse amplitude, we increase $\omega_r$ thereby modulating the number of oscillations the system undergoes, which demonstrates the coherence of the transfer process.

Figure 3c-d show the numerical solution to the master equation, which exhibit good agreement with the measured results. We plot these figures as a function of $\Omega_0{}^2$, where $\Omega_0$ is the peak amplitude of the time-varying Stark pulse (assumed to be a Gaussian pulse), again expressed as a classical Rabi frequency. Optimal transfer occurs at of $\Omega_0/2\pi = 107$ GHz, which corresponds to a peak Stark shift of 40 GHz. We note that there is a deviation in anti-correlation between the dot and cavity intensity at higher Stark powers, which is due to non-adiabatic corrections (see Supplement).

To further demonstrate the coherent control we perform a Ramsey-type experiment. In contrast to the previous results where the quantum dot and cavity were detuned, we perform the Ramsey measurement when the two systems are on-resonance. In this regime the two stationary states of the system are given by the dressed polaritons $\left| P_\pm \right\rangle = \left( \left| G,1 \right\rangle \pm \left| E,0 \right\rangle \right) / \sqrt{2}$. We choose this operating condition because both polaritons have equal decay rates, which leads to the best Ramsey interference contrast. We excite the polariton states simultaneously with a short 2 ps optical pulse, then let the system freely evolve for a fixed amount of time $\Delta\tau$ so that the polaritons dephase by $\phi = 2\pi\frac{\Delta\tau}{\tau_r}$. A Stark pulse subsequently mixes the two polariton states, creating a Ramsey interference effect. This experiment is analogous to Ramsey measurements performed on spin systems to demonstrate coherent control. Using the simplified model described by the Hamiltonian in Eq. 1, we show that the final occupation probabilities for the polaritons after the Stark pulse is given by $P_\pm = \frac{1}{2}(1 \mp \sin\phi \sin\omega_r\tau)$, where $\tau$ is once again the Stark pulse duration (see Supplement).    The oscillatory dependence on $\phi$ is the signature of Ramsey interference.

Figure 4a shows the measured spectrum as a function of $\Delta\tau$.  The two resonances observed in the spectrum for $\Delta\tau < 0$ correspond to the dressed-state polaritons.  For $\Delta\tau > 0$, the Stark pulse induces a clear energy transfer.  Figure 4b plots the emission intensity at the center wavelength of the lower and upper polaritons. The out-of-phase oscillation between the two polariton intensities is the signature of Ramsey interference.  These oscillations fall off after one period due to decay of the polariton states.

In conclusion, we have demonstrated picosecond optical control of vacuum Rabi oscillation and used it to coherently transfer excitation between a quantum dot and a cavity, as well as to perform coherent control of light-matter states.   Larger transfer contrasts could be attained by increasing the cavity-quantum dot coupling strength and reducing cavity losses using deterministic alignment[28] and better cavity designs[29,30].  Current state-of-the-art quantum dot cavity QED systems have already achieved $\frac{g}{\kappa} > 3$ [31].  Such coupling strengths may be sufficient to control higher order photon number states when combined with the control technique we demonstrate here.  Our results could ultimately provide a pathway towards gigahertz rate controlled synthesis of non-classical light at optical frequencies.

## Methods

### Device design and fabrication

The modes of the photonic molecule were calculated by a numerical finite-difference time-domain method using commercial software (Lumerical Inc.). The device design consisted of a two-dimensional array of air-holes in a triangular lattice with a radius of 70 nm and a period of 240 nm. Cavities were formed by removing three holes and shifting adjacent holes to optimize the quality factor[32].

Device fabrication was performed on an initial wafer consisting of a 160 nm GaAs membrane on top of a 1 μm thick AlGaAs sacrificial layer.  The GaAs membrane contained a single layer of InAs quantum dots at the center (with quantum dot density 30 μm$^{-2}$) embedded in gallium arsenide (GaAs) photonic crystal structures. Photonic crystals were fabricated using electron-beam lithography, followed by chlorine-based inductively coupled plasma etching and a chemical wet etch to remove the sacrificial layer.

**Continuous wave reflectivity measurement**

The fabricated sample was mounted in a liquid helium cryostat and optically excited from the out-of-plane direction using a broadband laser diode with emission between 900-1000 nm. We predominantly excited one of the two cavities in the molecule and isolated the emission from the same cavity using a spatial filter for all measurements performed. Due to strong hybridization of the two modes, the exact choice of cavity to excite does not matter, as both will give identical results. Signal reflected from the cavity was isolated using cross-polarization reflection spectroscopy and sent to a spectrometer with resolution of 0.02 nm. For the anti-crossing spectrum in Figure 1d, the temperature of the photonic crystal device was precisely tuned between 40 and 45 K.

**Picosecond optical pulse excitation**

Two time-synchronized Ti:Sapphire lasers were used as the excitation field and the Stark shift field. A phase-locked loop in the synchronization circuit controlled the delay between the two pulses. For data shown in Figures 2a-b and Figure 3 the excitation pulse was obtained by filtering 8 ps pulses from a Ti:Sapphire laser using a fiber Fabry-Perot filter to increase the pulse duration to approximately 15 ps. The filter bandwidth of the Fabry-Perot varied as it was tuned. Thus, the exact pulse duration varied from $10 - 22$ ps depending on the specific setting of the filter. The average excitation power was 20 nW. The Stark pulse was filtered to a 22 ps pulse duration using a free-space grating spectrometer (pulse duration measured using an autocorrelator), with an average excitation power of 120 nW. For the data in Figures 2c-d the excitation pulse was generated by filtering an 8 ps pulse using a fiber Fabry-Perot filter, and set to an average power of 40 nW. The Stark pulse was an 8 ps pulse directly generated by the laser with an average power of 150 nW. Results shown in Figure 4 were obtained using a 2 ps excitation pulse with an average power of 40 nW, and an 8 ps Stark pulse with average power of 130 nW, both obtained directly from the laser without filtering. For all experiments, the delay between the excitation and Stark pulse was calibrated using a high time resolution avalanche photodiode with temporal resolution of approximately 30 ps to determine the zero delay point.

**Acknowledgements:** The authors would like to acknowledge support from the Army Research Office Multidisciplinary University Research Initiative on Hybrid quantum interactions (grant number W911NF09104), the Physics Frontier Center at the Joint Quantum Institute, and the Office of Naval Research Applied Electromagnetics center. E. W. would like to acknowledge support from the National Science Foundation Faculty Early Career Development (CAREER) award (grant number ECCS. 0846494).

**Author Contributions:** R. B and T. C contributed equally. E.W conceived and designed the experiment. R. B and T.C designed and fabricated the device, conducted experiments and did analysis. K.R.C, E. W, T. C., and R. B performed theoretical simulations. R. B and E. W wrote the manuscript, with input from all authors. G. S. S grew the quantum dot wafer. E. W supervised the work.

# References

1       Hofheinz, M. *et al.* Generation of Fock states in a superconducting quantum circuit. *Nature* **454**, 310-314 (2008).

2       Hofheinz, M. *et al.* Synthesizing arbitrary quantum states in a superconducting resonator. *Nature* **459**, 546-549 (2009).

3       Atlasov, K. A., Rudra, A., Dwir, B. & Kapon, E. Large mode splitting and lasing in optimally coupled photonic-crystal microcavities. *Opt Express* **19**, 2619-2625 (2011).

4       Bayer, M. *et al.* Optical Modes in Photonic Molecules. *Phys. Rev. Lett.* **81**, 2582-2585 (1998).

5       Bose, R., Cai, T., Solomon, G. S. & Waks, E. All-optical tuning of a quantum dot in a coupled cavity system. *App. Phys. Lett.* **100**, 231107-231104 (2012).

6       Dousse, A. *et al.* Ultrabright source of entangled photon pairs. *Nature* **466**, 217-220 (2010).

7       Majumdar, A., Rundquist, A., Bajcsy, M. & Vučković, J. Cavity quantum electrodynamics with a single quantum dot coupled to a photonic molecule. *Phys. Rev. B* **86**, 045315 (2012).

8       Gehr, R. *et al.* Cavity-Based Single Atom Preparation and High-Fidelity Hyperfine State Readout. *Phys. Rev. Lett.* **104**, 203602 (2010).

9       Reithmaier, J. P. *et al.* Strong coupling in a single quantum dot-semiconductor microcavity system. *Nature* **432**, 197-200 (2004).

10      Yoshie, T. *et al.* Vacuum Rabi splitting with a single quantum dot in a photonic crystal nanocavity. *Nature* **432**, 200-203 (2004).

11      Englund, D. *et al.* Controlling cavity reflectivity with a single quantum dot. *Nature* **450**, 857-861 (2007).

12      Bochmann, J. *et al.* Fast Excitation and Photon Emission of a Single-Atom-Cavity System. *Phys. Rev. Lett.* **101**, 223601 (2008).

13      Englund, D. *et al.* Ultrafast Photon-Photon Interaction in a Strongly Coupled Quantum Dot-Cavity System. *Phys. Rev. Lett.* **108**, 093604 (2012).

14      Bose, R., Sridharan, D., Kim, H., Solomon, G. S. & Waks, E. Low-Photon-Number Optical Switching with a Single Quantum Dot Coupled to a Photonic Crystal Cavity. *Phys. Rev. Lett.* **108**, 227402 (2012).

15      Volz, T. *et al.* Ultrafast all-optical switching by single photons. *Nat Photon* **6**, 605-609 (2012).

16      Loo, V. *et al.* Optical Nonlinearity for Few-Photon Pulses on a Quantum Dot-Pillar Cavity Device. *Phys. Rev. Lett.* **109**, 166806 (2012).

17      Boozer, A. D., Boca, A., Miller, R., Northup, T. E. & Kimble, H. J. Reversible State Transfer between Light and a Single Trapped Atom. *Phys. Rev. Lett.* **98**, 193601 (2007).

18      Ritter, S. *et al.* An elementary quantum network of single atoms in optical cavities. *Nature* **484**, 195-200 (2012).

19      Kim, H., Bose, R., Shen, T. C., Solomon, G. S. & Waks, E. A quantum logic gate between a solid-state quantum bit and a photon. *Nat Photon* **7**, 373-377, doi:10.1038/nphoton.2013.48 (2013).

20      Reiserer, A., Ritter, S. & Rempe, G. Nondestructive Detection of an Optical Photon. *Science* **342**, 1349-1351 (2013).

21      Fernée, M. J., Rubinsztein-Dunlop, H. & Milburn, G. J. Improving single-photon sources with Stark tuning. *Phys. Rev. A* **75**, 043815 (2007).

22      Ates, S. *et al.* Non-resonant dot-cavity coupling and its potential for resonant single-quantum-dot spectroscopy. *Nat Photon* **3**, 724-728 (2009).

23      Fluegel, B. *et al.* Femtosecond Studies of Coherent Transients in Semiconductors. *Phys. Rev. Lett.* **59**, 2588-2591 (1987).

24      Sokoloff, J. P. *et al.* Transient oscillations in the vicinity of excitons and in the band of semiconductors. *Phys. Rev. B* **38**, 7615-7621 (1988).

25      Lange, C. *et al.* Ultrafast nonlinear optical response of photoexcited Ge/SiGe quantum wells: Evidence for a femtosecond transient population inversion. *Phys. Rev. B* **79**, 201306 (2009).

26    Guenther, T. *et al.* Coherent Nonlinear Optical Response of Single Quantum Dots Studied by Ultrafast Near-Field Spectroscopy. *Phys. Rev. Lett.* **89**, 057401 (2002).

27    Tan, S. M. A computational toolbox for quantum and atomic optics. *J. Opt. B*

28    Hennessy, K. *et al.* Quantum nature of a strongly coupled single quantum dot-cavity system. *Nature* **445**, 896-899 (2007).

29    Song, B.-S., Noda, S., Asano, T. & Akahane, Y. Ultra-high-Q photonic double-heterostructure nanocavity. *Nat Mater* **4**, 207-210 (2005).

30    Ota, Y. *et al.* Vacuum Rabi splitting with a single quantum dot embedded in a H1 photonic crystal nanocavity. *App. Phys. Lett.* **94**, - (2009).

31    Arakawa, Y., Iwamoto, S., Nomura, M., Tandaechanurat, A. & Ota, Y. Cavity Quantum Electrodynamics and Lasing Oscillation in Single Quantum Dot-Photonic Crystal Nanocavity Coupled Systems. *IEEE J. Sel. Top. Quant. Electron.* **18**, 1818-1829 (2012).

32    Akahane, Y., Asano, T., Song, B.-S. & Noda, S. High-Q photonic nanocavity in a two-dimensional photonic crystal. *Nature* **425**, 944-947 (2003).

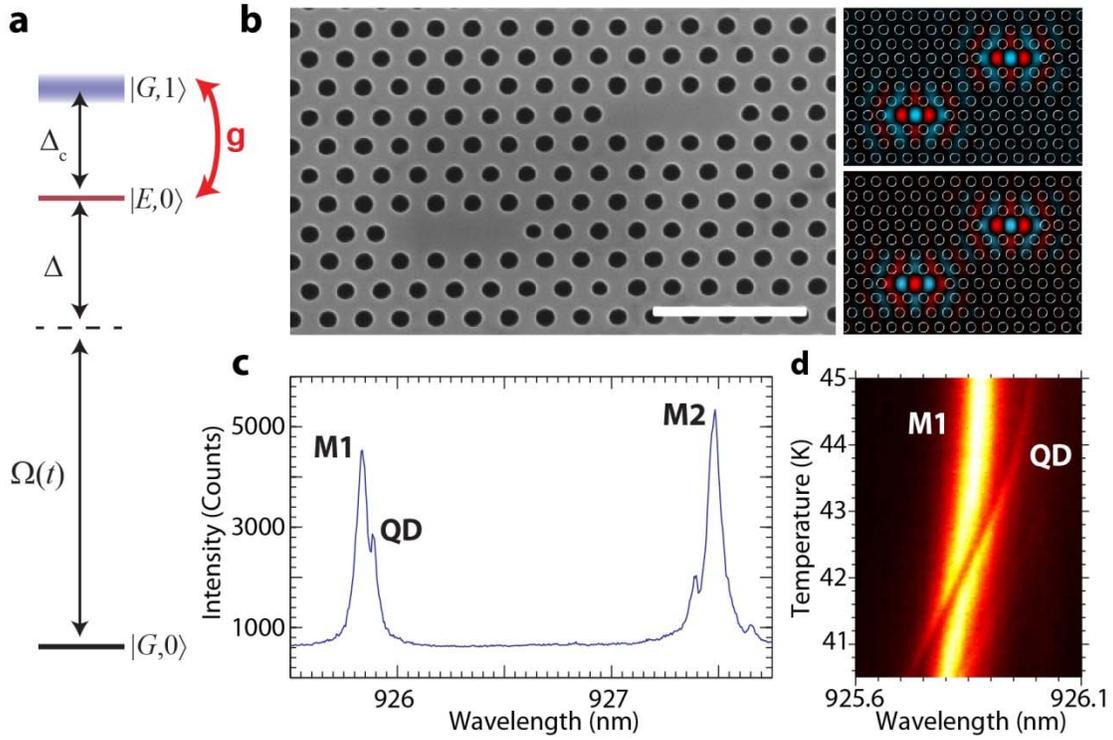

**Figure 1: Device Design and Characterization.** (a) Schematic of quantum level structure of a cavity-quantum dot system in the presence of a strong off-resonant pulse denoted by the classical Rabi frequency $\Omega(t)$. The pulse induces an AC Stark shift that optically tunes the quantum dot on resonance with the cavity. (b) SEM of fabricated photonic crystal molecule composed of two evanescently coupled photonic crystal cavities (Scale bar: 1 μm), along with finite-difference time-domain calculations showing the $\mathbf{E}_y$ component of the anti-symmetric (bottom) and symmetric (top) modes of the device. (c) Reflection spectrum of the photonic molecule, recorded at a temperature of 45 K, showing the two coupled cavity modes (denoted M1 and M2) and the quantum dot (labeled as QD). (d) Reflection spectrum around mode M1 as a function of temperature. The quantum dot tunes across the cavity mode, exhibiting an anti-crossing.

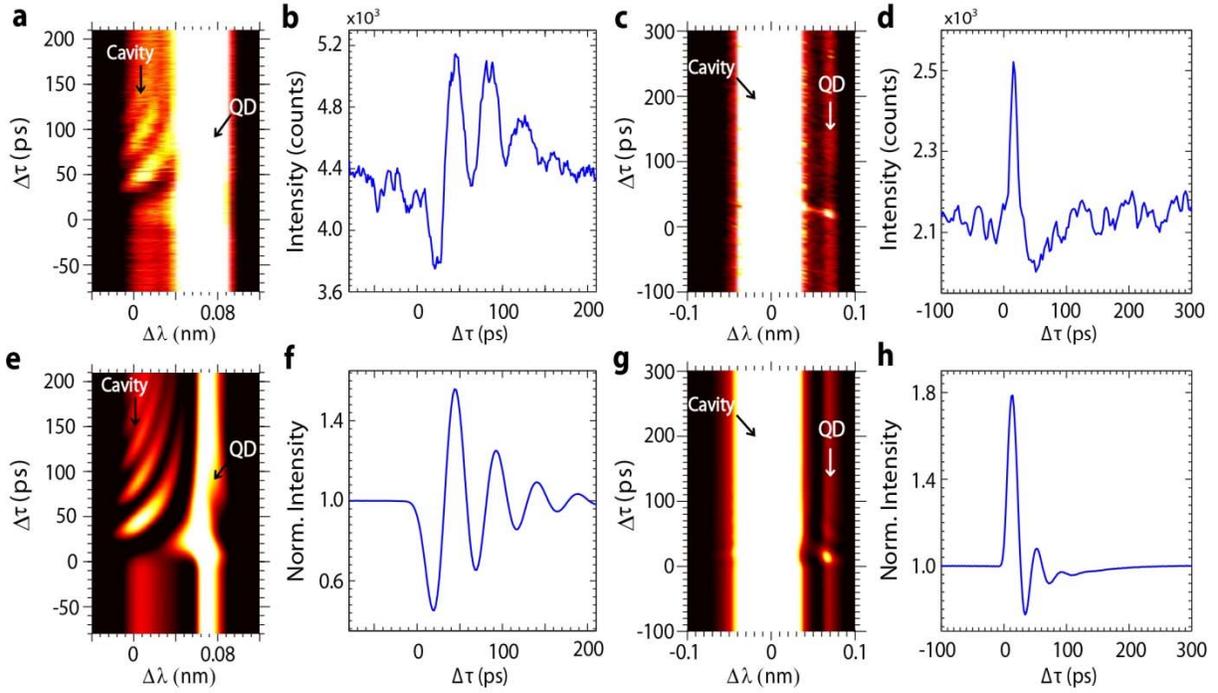

**Figure 2: Stark shift mediated energy transfer.** (a) Cavity spectrum as a function of delay $\Delta\tau$ when the excitation pulse excites the quantum dot resonance. (b) Intensity at $\Delta\lambda=0$, determined from Figure 2a, as a function of delay $\Delta\tau$. (c) Emission spectrum when the cavity is excited instead of the quantum dot. (d) Intensity at $\Delta\lambda= 0.07$ nm (quantum dot resonance) as a function of delay $\Delta\tau$ using data in figure 2c. (e)-(h) Calculation results using master equation for data corresponding to (a)-(d) respectively. In panels f and h, the intensities are normalized to the values at $\Delta\tau = $ -100 ps.

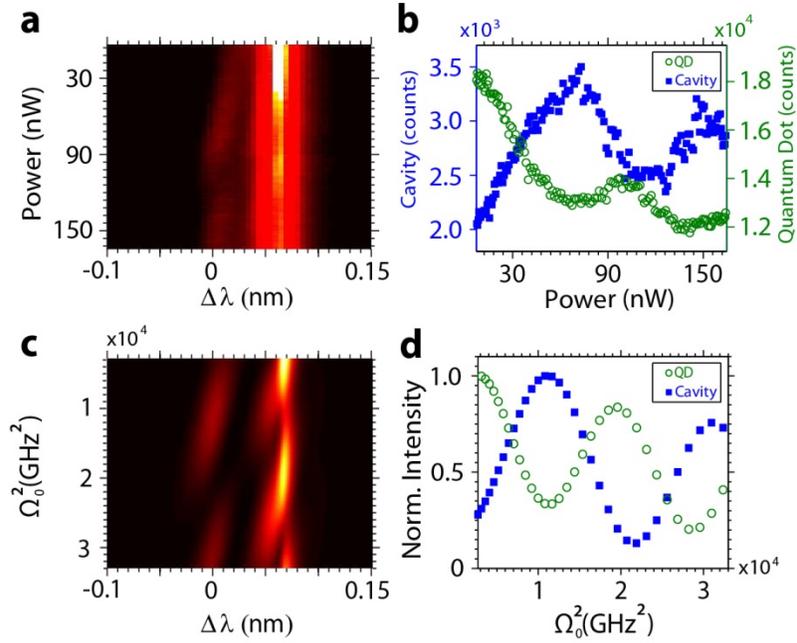

**Figure 3. Rabi oscillations.** (a) Measured reflections spectrum as a function of Stark laser power. (b) Emission intensity at the cavity resonance (blue squares) and at the quantum dot resonance (green circles), determined from the data in panel a. (c) Calculated spectrum as a function of Stark power. The Stark field is expressed as a classical Rabi frequency with peak amplitude $\Omega_0$. (d) Calculated emission intensity at cavity resonance ($\Delta\lambda = 0$) and the quantum dot resonance ($\Delta\lambda = 0.07$ nm). Intensities are normalized by their maximum value.

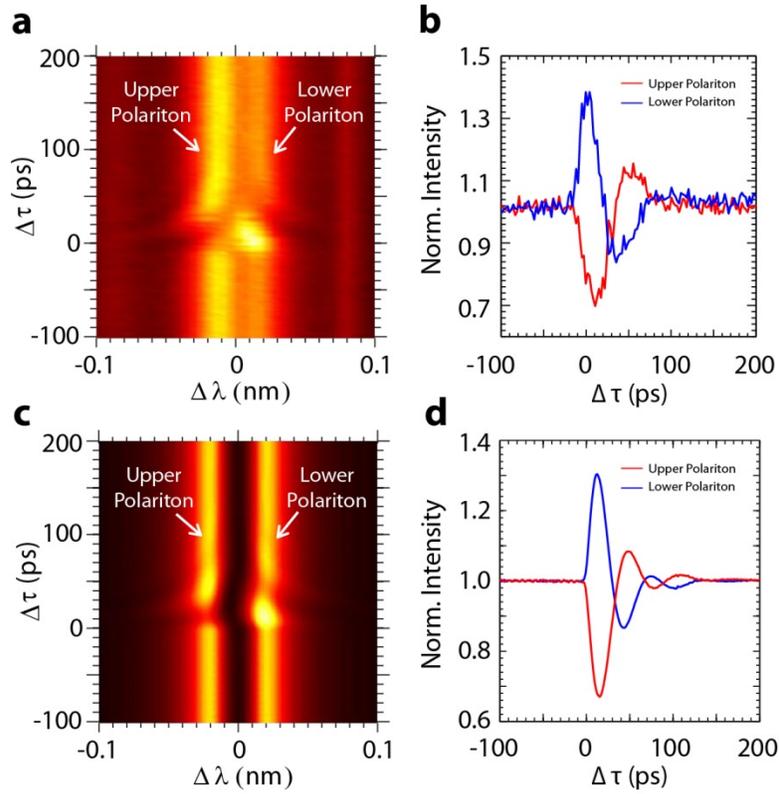

**Figure 4. Coherent control of polariton energy transfer.** (a) Emission spectrum as a function of delay $\Delta\tau$ when the quantum dot and cavity are resonant ($\Delta_C = 0$). (b) Emission intensity at the lower and upper polariton resonances as a function of delay between pump and probe. (c) Calculated cavity spectrum as a function of delay between the excitation and Stark shift pulse. (d) Calculated emission intensity at the lower polariton and upper polariton resonances. In panels b and d, intensities are normalized to the values at $\Delta\tau = -100$ ps.